\documentstyle{l-aa}
\begin{document}
\thesaurus{07(07.13.1; 07.09.1)}
\title{Alpha Cygnids - a possible July minor meteor shower}
\author{Arkadiusz Olech\inst{1,2}, ~Marcin Gajos\inst{1,2}
\and Micha{\l} Jurek\inst{2}}
\offprints{Arkadiusz Olech, e-mail: olech@sirius.astrouw.edu.pl}
\institute{Warsaw University Observatory, Al. Ujazdowskie 4, 00-478
Warszawa, Poland
\and
Comets and Meteors Workshop, ul. Sokolicz 3/59, 01-508 Warszawa, Poland}
 
\date{Received June ......, 1998, Accepted ...................., 1998}
\maketitle
\begin{abstract}
 
We present a comprehensive study of a possible $\alpha$-Cygnid meteor
shower. Based on visual and telescopic observations made by Polish observers 
we estimate basic parameters of the stream.
Activity of $\alpha$-Cygnids lasts from around June 30 to July 31
with clear maximum near July 18 (solar longitude $\lambda_\odot=116.5^\circ$).
Maximal Zenithal Hourly Rates (ZHRs) are equal to $3.6\pm1.2$. The structure of
the radiant analyzed by {\sc radiant} software is most compact for geocentric
velocity of the events equal to 41 km/s, and for the drift of the
radiant (in units $^\circ$/day) equal to 
$\Delta\alpha=+0.6$ $\Delta\delta=+0.2$. Center of
the radiant for the moment of maximum is $\alpha=302.5^\circ$
$\delta=+46.3^\circ$.We derive population index $r$ equal to
$2.55\pm0.14$ from magnitude distribution of 738 possible members of
the stream. Comparing the
velocity distributions of 754 possible $\alpha$-Cygnids and 4339
sporadic events by $\chi^2$ and Kolmogorov-Smirnov tests we conclude
that both distributions are different with probability very close to 1.0.

Telescopic observations strictly confirm the results obtained from
visual observations. The smallest values of $\chi^2$ parameter we
obtained for the geocentric velocity equal to 40 km/s and for the drift of
the radiant (in units $^\circ$/day) 
equal to $\Delta\alpha=+0.6$ $\Delta\delta=+0.2$.  The
center of the radiant for moment of maximum is $\alpha=304.9^\circ$
$\delta=+46.2^\circ$.

In spite of making many photographic exposures we still have no
photographic or video confirmation of the existence of this stream.

\end{abstract}
 
\section{Introduction}

The first informations about meteors from radiant near $\alpha$ Cygni
come from W.F. Denning (Denning 1919). In years 1885--1918 he observed
50 meteors radiating from the close vicinity of Deneb. He did not know
the activity period of the stream so he noted meteors during the whole
year. It is clearly visible from his Table 2 that majority of meteors
were noted during July nights. Almost all events observed in this month 
were classified as rapid.

During next years one can find the different parameters describing
$\alpha$-Cygnids stream in the astronomical literature. Polish meteor
publications were giving activity period between June 16th and July
31st. No clear maximum of activity was found (Kosinski 1990).

Photographic data are also poor. Only one possible member was captured in
Dushanbe on 1961 July 12. The radiant of this event was
$\alpha=304.5^\circ$ ~ $\delta=+49.7^\circ$ and geocentric velocity
$V_\infty=41.0$~km/s (Babadzhanov \& Kramer 1965).

\begin{center}
\begin{table*}
\caption[ ]{List of visual CMW observers}
\begin{flushleft}
\begin{tabular}{||l||r|r|r||r|r|r||r|r|r||}
\hline
\hline
 & \multicolumn{3}{|c||}{1995} & \multicolumn{3}{|c||}{1996} &
\multicolumn{3}{|c||}{1997}\\
\cline{2-10}
Observer & $t_{eff}$ & $N_{\alpha}$ & $N_{sp}$ & $t_{eff}$ &
$N_{\alpha}$ & $N_{sp}$ & $t_{eff}$ & $N_{\alpha}$ & $N_{sp}$ \\
\hline
\hline
Maciej Reszelski & $56^h51^m$ & 40 & 278 & $27^h22^m$ & 35 & 181 &
$11^h58^m$ & 18 & 114 \\
Arkadiusz Olech  & $32^h45^m$ & 35 & 257 & $16^h55^m$ & 19 & 129 &
$31^h31^m$ & 39 & 234 \\
Tomasz Fajfer & - & - & - &  $31^h30^m$ & 33 & 134 & $39^h30^m$ & 52 &
341 \\
Konrad Szaruga & - & - & - & $11^h28^m$ & 6 & 84 & $50^h16^m$ & 85 & 257
\\
Marcin Gajos & $26^h00^m$ & 12 & 144 & $6^h18^m$ & 4 & 38 & $10^h40^m$ &
11 & 83 \\
Krzysztof Socha & $17^h00^m$ & 21 & 103 & $6^h54^m$ & 4 & 38 &
$18^h16^m$ & 12 & 142 \\
Maciej Kwinta & $4^h00^m$ & 0 & 13 & $8^h10^m$ & 0 & 13 & $26^h50^m$ & 9
& 157 \\
Krzysztof Wtorek & $5^h42^m$ & 5 & 32 & $10^h00^m$ & 14 & 67 &
$12^h55^m$ & 12 & 45 \\
{\L}ukasz Sanocki & $14^h30^m$ & 17 & 62 & $5^h46^m$ & 8 & 21 &
$8^h04^m$ & 5 & 27 \\
Tomasz Dziubi\'nski & $15^h47^m$ & 16 & 35 & $3^h30^m$ & 5 & 20 &
$6^h00^m$ & 3 & 25 \\
Jaros{\l}aw Dygos & - & - & - & - & - & - & $24^h30^m$ & 16 & 157 \\
Gracjan Maciejewski & - & - & - & - & - & - & $21^h15^m$ & 8 & 91 \\
Andrzej Skoczewski & - & - & - & - & - & - & $20^h23^m$ & 28 & 85 \\
Artur Szaruga & - & - & - & - & - & - & $19^h55^m$ & 14 & 71 \\
Wojciech Jonderko & - & - & - & $4^h15^m$ & 0 & 11 & $15^h24^m$ & 7 & 67
\\
Marcin Konopka & - & - & - & - & - & - & $18^h39^m$ & 17 & 181 \\
Micha{\l} Jurek & - & - & - & $9^h31^m$ & 9 & 80 & $7^h39^m$ & 9 & 53 \\
Tomasz \.Zywczak & - & - & - & - & - & - & $16^h38^m$ & 10 & 43 \\
Maria Wo\'zniak & $13^h45^m$ & 18 & 43  & - & - & - & - & - & - \\
Tomasz Ramza & - & - & - & $6^h30^m$ & 5 & 49 & $5^h59^m$ & 1 & 32 \\
{\L}ukasz Pospieszny & - & - & - & $9^h44^m$ & 4 & 62 & $3^h31^m$ & 1 &
18 \\
Piotr Grzywacz & $11^h30^m$ & 21 & 55  & - & - & - & - & - & - \\
Robert Szczerba  & - & - & - & $5^h00^m$ & 6 & 21 & $6^h20^m$ & 1 & 100
\\
Krzysztof Kami\'nski  & - & - & - & - & - & - & $8^h47^m$ & 6 & 60 \\
{\L}ukasz Raurowicz & - & - & - & $1^h07^m$ & 2 & 3 & $6^h16^m$ & 6 & 44
\\
Tadeusz Sobczak & - & - & - & - & - & - & $6^h10^m$ & 20 & 46 \\
Rafa{\l} Kopacki & - & - & - & $5^h30^m$ & 8 & 45 & - & - & - \\
Micha{\l} Marek & - & - & - & $4^h00^m$ & 6 & 5 & $0^h30^m$ & 1 & 4 \\
El\.zbieta Brembor & $4^h00^m$ & 8 & 5  & - & - & - & - & - & - \\
Ireneusz S{\l}awi\'nski & - & - & - & $3^h00^m$ & 0 & 12 & - & - & - \\
Krzysztof Gdula & - & - & - & $2^h30^m$ & 2 & 11  & - & - & - \\
Marek Piotrowski & - & - & - & - & - & - & $2^h25^m$ & 2 & 14 \\
Pawe{\l} Trybus & - & - & - & - & - & - & $2^h00^m$ & 0 & 10 \\
Pawe{\l} Musialski & - & - & - & $1^h30$ & 1 & 7 & - & - & - \\
Pawe{\l} Gembara & $1^h00^m$ & 0 & 10 & - & - & - & - & - & - \\
\hline
Total & $202^h50^m$ & 193 & 1037 & $180^h30^m$ & 171 & 1031 &
$402^h21^m$ & 393 & 2501 \\
\hline
\hline
\end{tabular}
\end{flushleft}
\end{table*}
\end{center}
 
\begin{center}
\begin{table*}
\caption[ ]{List of telescopic CMW observers}
\begin{flushleft}
\begin{tabular}{||l||r|r|r||r|r|r||r|r|r||}
\hline
\hline
 & \multicolumn{3}{|c||}{1996} & \multicolumn{3}{|c||}{1997} &
\multicolumn{3}{|c||}{Total}\\
\cline{2-10}
Observer & $t_{eff}$ & $N_{\alpha}$ & $N_{sp}$ & $t_{eff}$ &
$N_{\alpha}$ & $N_{sp}$ & $t_{eff}$ & $N_{\alpha}$ & $N_{sp}$ \\
\hline
\hline
Tomasz Dziubi\'nski & $4^h51^m$ & 7 & 20 & $4^h40^m$ & 5 & 31 &
$9^h31^m$ & 12 & 51 \\
Micha{\l} Jurek & $1^h56^m$ & 1 & 3 & $3^h31^m$ & 1 & 13 & $5^h27^m$ & 2
& 16 \\
Konrad Szaruga & $1^h21^m$ & 3 & 5 & $3^h58^m$ & 6 & 34 & $5^h19^m$ & 9
& 39 \\
Marcin Gajos & $1^h49^m$ & 1 & 10 & $2^h42^m$ & 4 & 17 & $4^h31^m$ & 5 &
27 \\
Jaros{\l}aw Dygos & - & - & - & $4^h09^m$ & 1 & 10 & $4^h09^m$ & 1 & 10
\\
Tomasz Fajfer & $2^05^m$ & 2 & 9 &  $2^h00^m$ & 2 & 9 & $4^h05^m$ & 4 &
18 \\
Marcin Konopka & - & - & - & $3^h31^m$ & 1 & 13 & $3^h31^m$ & 1 & 13 \\
Wojciech Jonderko & - & - & - & $1^h23^m$ & 0 & 0 & $1^h23^m$ & 0 & 0 \\
{\L}ukasz Pospieszny & - & - & - & $1^h11^m$ & 1 & 3 & $1^h11^m$ & 1 & 3
\\
Rafa{\l} Kopacki & $1^h00^m$ & 0 & 3 & - & - & - & $1^h00^m$ & 0 & 3 \\
Krzysztof Wtorek & $1^h00^m$ & 0 & 3 & - & - & - & $1^h00^m$ & 0 & 3 \\
Maciej Reszelski & - & - & - & $0^h59^m$ & 4 & 6 & $0^h59^m$ & 4 & 6 \\
Micha{\l} Kopczak & $0^h53^m$ & 1 & 1 & - & - & - & $0^h53^m$ & 1 & 1 \\
Andrzej Skoczewski & - & - & - & $0^h44^m$ & 0 & 4 & $0^h44^m$ & 0 & 4
\\
\hline
Total & $14^h55^m$ & 15 & 54 & $28^h32^m$ & 26 & 139 & $43^h27^m$ & 41 &
193 \\
\hline
\hline
\end{tabular}
\end{flushleft}
\end{table*}
\end{center}

In the comprehensive work undertaken by Dutch Meteor Society (DMS) and North
Australian Planetary Observers -- Meteor Section (NAPO--MS) in years
1981-1991 and described in detail by Jenniskens (1994) one can read
about weak stream called $o$-Cygnids. During 98 hours of
effective time of observations 8 observers noted 72 possible members of
that stream. From this data Jenniskens (1994) estimated the following
parameters of the stream:
\smallskip

\noindent $\bullet$ equatorial coordinates of the radiant during the
maximum of activity: $\alpha=305^\circ$ ~ $\delta=+47^\circ$,

\noindent $\bullet$ drift of the radiant (in units $^\circ$/day): 
$\Delta\alpha=+0.6$ ~ $\Delta\delta=+0.2$, 

\noindent $\bullet$ maximum of activity:
$\lambda_{\odot(1950.0)}=116.0^\circ\pm0.5^\circ$, 

\noindent $\bullet$ population index $r=2.7$, where $r$ is defined as:
\begin{equation}
r={{\Phi (m+1)}\over{\Phi (m)}}
\end{equation}
 
\noindent where 
 
\begin{equation}
\Phi (m) = \sum^{m}_{-\infty} N(m)
\end{equation}
 
\noindent and $N(m)$ is the
number of meteors with magnitude $m$ corrected for probabilities of
perception given by Koschack and Rendtel (1990),

\noindent $\bullet$ Maximal Zenithal Hourly Rates (ZHRs) are equal to
$2.5\pm0.8$, where ZHR is defined as:
\begin{equation}
ZHR={{N_h\cdot r^{(6.5-LM)}}\over{{(\sin H)}^\gamma}}
\end{equation}
 
\noindent where $N_h$ is the observed number of meteors per hour
(corrected
for clouds coverage), $LM$ is the limiting magnitude in the field of
view, $H$ is an altitude of the radiant of the stream, and $\gamma$ is a
zenith exponent factor,

\noindent $\bullet$ geocentric velocity: $V_\infty=37$~km/s.

\smallskip

From the Fig. 11 in paper by Jenniskens (1994) it is clearly visible
that activity of $o$-Cygnids lasts from $\lambda_\odot\approx105^\circ$
to $\lambda_\odot\approx127^\circ$ i.e. from around July 5th to July
27th.

\section{Observations} 

\subsection{Visual observations}

July is a good month for meteor observers in Poland. Polish observers
associated in Comets and Meteors Workshop (CMW) are mostly young people
studying at secondary schools or universities. July is the first month
of summer holidays. The warm nights, a lot of free time, good weather
conditions strongly encourage to observations. Finally about 60--80\% of
whole year observations made by CMW members are collected in July and
August. 

All above facts and rapidly growing interest in meteor observations in
Poland during the last few years allowed us to investigate the activity of the
stream called $\alpha$-Cygnids or $o$-Cygnids. Each July in years
1995-1997 many meteor observers watched the sky using visual, telescopic
and photographic techniques. In order to obtain most reliable
results we had to remove observations made incorrectly or in poor
conditions. Using our standard methods (Olech \& Wo\'zniak 1996, Olech
1997) we required that:
\smallskip

\noindent $\bullet$ mean limiting magnitude (LM) in the field of view
should be at least 5.0 mag,

\noindent $\bullet$ cloud coverage correction factor $F$ should be
smaller than 1.7,

\noindent $\bullet$ the radiant of the stream should be above $20^\circ$
over horizon (in Polish latitudes during whole July this condition is always
satisfied),

\noindent $\bullet$ center of the field of view should be at altitude
higher than $40^\circ$.
\smallskip

Finally we obtained $785^h41^m$ of effective time of visual observations. During
this time a group of 35 CMW members noted 757 possible meteors from
$\alpha$-Cygnid stream and 4569 sporadic meteors. The more detailed
statistics with names of CMW observers is given in Table 1.

\subsection{Telescopic observations}

Telescopic observations present a very useful tool for meteor investigators.
Meteors are very often plotted with larger accuracy than in case of visual
observations. It gives the possibility to study the structure and drift
of the radiant. We also obtain informations about magnitude
distribution for fainter events. The main problem with telescopic
observations is that this kind of watching meteors requires good
equipment (preferably binoculars with large field of view mounted on tripod),
experienced observers and a lot of patience.

Fortunately July is usually the time in which we organize
an Astronomical Camp of CMW, which
takes place at the Observational Station of Warsaw University Observatory in
Ostrowik. The number of participants is always around 15, so we organize
two four persons groups observing visually, one or two persons working with a few
cameras pointed at different directions, and three--four persons
observing telescopicaly different fields located $10^\circ-40^\circ$ from
supposed radiant
of $\alpha$-Cygnids. We used mostly $7\times50$,
$10\times50$ and $20\times60$ binoculars and Russian AT-1 refractors
($5\times50$, field of view as large as $11^\circ$). For plotting meteors
we use $A$-type maps of International Meteor Organization (IMO) or {\it
Uranometria 2000.0} charts. Of course other observers which do not
participate in the camp observe $\alpha$-Cygnids both visually and
telescopicaly at their locations.

Finally 14 our observers obtained $43^h27^m$ of telescopic observations with 234
meteors detected. The number of possible $\alpha$-Cygnids equals to 41. Table 2
summarizes our telescopic observations.

\begin{center}
\begin{table*}
\caption[ ]{Magnitude distribution for 1995-1997 $\alpha$-Cygnids}
\begin{flushleft}
\begin{tabular}{|l|ccccccccccccc|c|}
\hline
\hline   
Year & $\leq$-5 & -4 & -3 & -2 & -1 & 0 & 1 & 2 & 3 & 4 & 5 & 6 & 7 &
Tot. \\
\hline
\hline
1995 & 0 & 0 & 3.5 & 1.5 & 4 & 5 & 32.5 & 72.5 & 35 & 23 & 14 & 2 & 0 &
193 \\
1996 & 0 & 0 & 0.5 & 3.5 & 4 & 7.5 & 13 & 25 & 45.5 & 42 & 25 & 5 & 0 &
171 \\
1997 & 0 & 3 & 0.5 & 2.5 & 6 & 18.5 & 32 & 65.5 & 93 & 81.5 & 53 & 18.5
& 0 & 374 \\
\hline
Tot. & 0 & 3 & 4.5 & 7.5 & 14 & 31 & 77.5 & 163 & 173.5 & 146.5 & 92 &
25.5 & 0 & 738 \\
\hline
\hline
\end{tabular}
\end{flushleft}
\end{table*}
\end{center}
\vspace{1cm}
 
\begin{center}
\begin{table*} 
\caption[ ]{Magnitude distribution for 1995-1997 sporadics}
\begin{flushleft}
\begin{tabular}{|l|ccccccccccccc|c|}
\hline
\hline
Year & $\leq$-5 & -4 & -3 & -2 & -1 & 0 & 1 & 2 & 3 & 4 & 5 & 6 & 7 &
Tot. \\
\hline
\hline
1995 & 0 & 1 & 4.5 & 6 & 16 & 62.5 & 139.5 & 252.5 & 268.5 & 173 & 87.5
& 16 & 0 & 1027 \\
1996 & 0 & 0 & 4.5 & 10 & 32 & 72.5 & 123.5 & 197.5 & 239 & 202.5 & 132
& 17.5 & 0 & 1031 \\
1997 & 5 & 16 & 11 & 25 & 67.5 & 123 & 274.5 & 405.5 & 547.5 & 577.5 &
348.5 & 86.5 & 0.5 & 2488 \\
\hline
Tot. & 5 & 17 & 20 & 41 & 115.5 & 258 & 537.5 & 855.5 & 1055 & 953 & 568
& 120 & 0.5 & 4546 \\ 
\hline
\hline
\end{tabular}
\end{flushleft}
\end{table*}
\end{center}
\vspace{1cm}

\begin{center}
\begin{table*}
\caption[ ]{Magnitude distribution for 1996-1997 telescopic $\alpha$-Cygnids}
\begin{flushleft}
\begin{tabular}{|l|cccccc|c|}
\hline
\hline   
Year & 4 & 5 & 6 & 7 & 8 & 9 & Tot. \\
\hline
\hline
1996 & 0.5 & 0.5 & 6 & 6.5 & 1.5 & 0 & 15 \\
1997 &  0  &  2  & 3 & 8.5 & 7.5 & 5 & 26 \\
\hline
Tot. & 0.5 & 2.5 & 9 & 15 & 9 & 5 & 41 \\
\hline
\hline
\end{tabular}
\end{flushleft}
\end{table*}
\end{center}

\begin{center}
\begin{table*}
\caption[ ]{Magnitude distribution for 1996-1997 telescopic sporadics}
\begin{flushleft}
\begin{tabular}{|l|cccccccccc|c|}
\hline
\hline
Year & 1 & 2 & 3 & 4 & 5 & 6 & 7 & 8 & 9 & 10 & Tot. \\
\hline
\hline
1996 & 0 & 0.5 &  3  & 1 & 6.5 &  8.5 & 24.5 &  9   &  1 & 0 & 54 \\
1997 & 1 & 0.5 & 2.5 & 6 & 5.5 & 18.5 & 51.5 & 39.5 & 12 & 2 & 139 \\
\hline
Tot. & 1 &  1  & 5.5 & 7 & 12  & 27   & 76  & 48.5 & 13 & 2 & 193 \\
\hline
\hline
\end{tabular}
\end{flushleft}
\end{table*}
\end{center}

\section{Results}

\subsection{Radiant of $\alpha$-Cygnids}

During July of 1995, 1996 and 1997 CMW observers plotted on gnomonic star maps
2748 paths of meteor events. For each of them the angular velocity was
estimated. We used 0--5 scale with 0 corresponding to stationary meteor,
1 to very slow event, 2 to slow, 3 to medium, 4 to fast and 5 to very
fast meteor. Equatorial coordinates of the begins and ends of these
events and their velocities were put into the {\sc radiant} software
(Arlt 1992). This software as an input also requires the geocentric
velocity of the meteors $V_\infty$ and the daily drift of the radiant.
Changing both these values we can obtain different density distributions
of the probability area near suspected radiant. Choosing the best
distribution (this one with smallest $\chi^2$ parameter) we are able to
estimate the values of $V_\infty$ and the daily drift. The systematic
errors play a role, which are difficult to handle and estimate of the
accuracy of the obtained value of $V_\infty$ is difficult but the errors
are at minimum $\pm5$~km/s. For more details see Arlt (1993).

Before analyzing our sample we decided to analyze also the meteors observed by
Denning (1919). However we selected only meteors observed by him during
July nights. Number of these events accounted to 20. We performed 
our calculation using parameters of the stream given by Jenniskens (1994) i.e.
$V_\infty=37$~km/s, $\lambda_{\odot{\rm (max)}}=116^\circ$, $\Delta\alpha=+0.6$ and
$\Delta\delta=+0.2$. Results as probability function distribution of the 
presence of radiant are presented in Fig. 1. The best fit of the two
dimensional Gaussian surface to the density of probability map gives
coordinates of the radiant equal to $\alpha=312.4^\circ$ and $\delta=+48.4^\circ$. The
accuracy of this estimate is certainly low due to small number of
events observed by Denning (1919).

Fortunately, the sample collected by CMW observers in years 1995-1997 is
significantly larger. It allows us to derive a few valuable conclusions.
First we calculate our sample (2748 meteors including possible member of
the stream, sporadics and meteors from other showers) using parameters given by Jenniskens
(1994). During calculation we remove meteors observed at distance
larger than $85^\circ$ from the radiant of the stream. The prominence of the
$\alpha$-Cygnid radiant on the resulting picture is striking. The best fit gives
coordinates of the radiant as $\alpha=302.0^\circ$ and $\delta=+46.1^\circ$.
Nevertheless we obtain better results i.e.
more compact shape of the radiant using geocentric velocity
$V_\infty=41$~km/s and the drift of the radiant  $\Delta\alpha=+0.6$, 
$\Delta\delta=+0.2$. The resulting radiant picture for the above
parameters is displayed in Fig. 2. The final coordinates of the radiant
of $\alpha$-Cygnid stream are $\alpha=302.5^\circ$ and
$\delta=+46.3^\circ$, which do not differ significantly from coordinates obtained for
parameters given by Jenniskens (1994).

We also used the {\sc radiant} software for the analysis of the paths of our
telescopic meteors. Our sample contains 234 meteors with known paths and
velocities. The resulting density distribution from
telescopic observations is displayed in Fig. 3. The best fit (with
smallest $\chi^2$ value) is obtained for the following parameters:
geocentric velocity $V_{\infty}=40$~km/s, the daily drift of the radiant
$\Delta\alpha=+0.6^\circ$ and $\Delta\delta=+0.2^\circ$. The
coordinates of the center of the radiant are $\alpha=304.9^\circ$ and
$\delta=+46.2^\circ$. One can see that the position of the radiant
obtained from
telescopic observations differ from the position obtained from visual
data by only $1.7^\circ$. Taking into account that radii of the majority
of radiants vary between $2^\circ$ and $7^\circ$ both our results are
strictly consistent. It is also clear that our parameters are in very
good agreement with the data of the one photographed meteor (Babadzhanov \& Kramer 1965).

\subsection{Population index r}

In years 1995-1997 the CMW observers made as many as 738 and 4546
estimates of the brightness of meteor events from $\alpha$-Cygnids
and sporadics, respectively. The distribution of this quantity is
presented in Tables 3 and 4.

Such a large amount of magnitude estimates for $\alpha$-Cygnids
encouraged us to compute the value of the population index $r$ defined
in equation (1). We obtained $r=2.55\pm0.14$ which is a typical value among
meteor streams. Jenniskens (1994) obtained similar result with $r$ equal
to 2.7. The population index obtained from magnitudes of our 4546
sporadics is equal to $r=2.61\pm0.23$.

Also the telescopic observers estimated the magnitudes of meteor
events. The magnitude distributions for 1996 and 1997 $\alpha$-Cygnids
and sporadics are presented in Tables 5 and 6.

\subsection{Activity profile}

Knowing the value of $r$ we can compute ZHR using
the formula given in (3). According to the results of  Koschack
(1994) and Bellot (1995) who showed that for visual observations with
radiant altitudes higher than $20^\circ$ the zenith exponent factor 
$\gamma\approx1.0$, we adopted $\gamma=1.0$.

The resulting activity profile of $\alpha$-Cygnids in years 1995--1997
is exhibited in Fig. 4. The activity of the stream lasts from
$\lambda_\odot\approx100^\circ$ (June 30) to
$\lambda_\odot\approx130^\circ$ (July 31). It seems to be slightly wider
than the result of Jenniskens (1994) who noted meteors from $\alpha$-Cygnid
stream in interval $\lambda_\odot=105-127^\circ$. The accuracy of the ZHR
estimates by Jenniskens (1994) was low due to the small number of his
observations, therefore we prefer our result. 

Our Fig. 4 one exhibits a clear maximum of activity at
$\lambda_\odot\approx116.5^\circ$ with ${\rm ZHR}=3.6\pm1.2$. The error of
this estimate is large but points in the
vicinity of the maximum have smaller errors and their moments and ZHRs
are $\lambda_\odot=114.5^\circ$ with ${\rm ZHR}=2.9\pm0.4$ and 
$\lambda_\odot=118.5^\circ$ with ${\rm ZHR}=3.1\pm0.6$

The moment of the maximum and its ZHR is in very good agreement with
result of Jenniskens (1994) who obtained 
$\lambda_{\odot{\rm (max)}}=116.0\pm0.5^\circ$ with ${\rm ZHR_{max}}=2.5\pm0.8$.

Jenniskens (1994) found also that the activity profiles of meteor streams
are well represented by the following equation:

\begin{equation}
{\rm ZHR} = {\rm ZHR_{max}}\cdot 10^{-B\cdot|\lambda_\odot -
\lambda_{\odot max}|}
\end{equation}

For $\alpha$-Cygnids he found $B=0.13\pm0.03$. As we have already written our activity
profile is much broader and thus our value of $B$ is smaller and equal to $0.045\pm0.005$. 
In Fig. 4 the solid line represents fit given in equation (4) with ${\rm
ZHR_{max}}=3.6$, $\lambda_{\odot max}=116.5^\circ$ and $B=0.045$.

\subsection{Velocity distribution}

The CMW observers estimated also the angular velocity of the events.
The 0--5 scale (defined in {\it Sec. 3.1} of this paper)
was used. Finally we obtained 754 estimates of the angular velocity for
$\alpha$-Cygnids and 4339 estimates for sporadics. The velocity
distribution from visual observations is presented in Tables 7--8.

\begin{center}
\begin{table}
\caption[ ]{Velocity distribution for 1995-1997 $\alpha$-Cygnids.} 
\begin{flushleft}
\begin{tabular}{|l|cccccc|c|}
\hline
\hline
Year & 0 & 1 & 2 & 3 & 4 & 5 & Tot. \\
\hline
\hline
1995 & 0 & 0 & 6 & 63 & 96 & 28 & 193 \\
1996 & 0 & 0 & 9 & 44 & 97 & 20 & 170 \\
1997 & 4 & 2 & 22 & 166 & 172 & 25 & 391 \\
\hline
Tot. & 4 & 2 & 37 & 273 & 365 & 73 & 754 \\
\hline
\hline
\end{tabular} 
\end{flushleft}
\end{table}
\end{center}

\begin{center}
\begin{table}
\caption[ ]{Velocity distribution for 1995-1997 sporadics}
\begin{flushleft}
\begin{tabular}{|l|cccccc|r|}
\hline
\hline
Year & 0 & 1 & 2 & 3 & 4 & 5 & Tot. \\
\hline
\hline
1995 & 6 & 17 & 68 & 247 & 418 & 171 & 927 \\
1996 & 4 & 22 & 86 & 243 & 433 & 233 & 1021 \\
1997 & 42 & 31 & 202 & 710 & 938 & 468 & 2391 \\
\hline
Tot. & 52 & 70 & 356 & 1200 & 1789 & 872 & 4339 \\
\hline
\hline
\end{tabular} 
\end{flushleft}
\end{table}
\end{center}

We used the above distributions to find another proof for existence of the
$\alpha$-Cygnid stream. We compared empirical velocity distributions of
$\alpha$-Cygnids and sporadics using Kolmogorov--Smirnov and $\chi^2$
tests. We obtained that with the probability larger than 0.999 both
distributions are different. Such a large probability is certainly
caused by the clear enhancement of meteors with velocity 3 and 4 in
$\alpha$-Cygnid velocity distribution. This result is also with good
agreement with the value of geocentric velocity obtained from {\sc radiant}
analysis of our visual and telescopic data. The meteors with velocity
$V_\infty=40-41$~km/s given by {\sc radiant} software at mean distance
from the radiant of the stream appear mainly with velocities 3 and 4 in 0--5 scale.

The velocity of meteor events was also estimated by our telescopic
observers. They used $A$--$F$ scale with $A$ corresponding to the
angular velocity $2^\circ/sec$ and $F$ to over $25^\circ/sec$. Finally we
obtained 41 estimates of the angular velocity for telescopic $\alpha$-Cygnids and 192
velocity estimates for telescopic sporadics. Both distributions are
presented in Tables 9--10.

For telescopic observations the distance from the radiant is generally
well defined. Usually it is worthwhile to analyze the mean angular
velocity as a function of distance from the radiant. Unfortunately due
to the small number of our telescopic $\alpha$-Cygnids which were
observed in as many as 10 fields such an analysis is impossible yet.

\begin{center}
\begin{table} 
\caption[ ]{Velocity distribution for 1996-1997 telescopic
$\alpha$-Cygnids}
\begin{flushleft}
\begin{tabular}{|l|ccccccc|c|}
\hline
\hline
Year & 0 & A & B & C & D & E & F & Tot. \\
\hline
\hline
1996 & 0 & 0 & 1 & 6 &  6 & 2 & 0 & 15 \\
1997 & 0 & 0 & 4 & 5 & 11 & 4 & 2 & 26 \\
\hline
Tot. & 0 & 0 & 5 & 11 & 17 & 6 & 2 & 41 \\
\hline
\hline
\end{tabular}
\end{flushleft}
\end{table}
\end{center}

\begin{center}
\begin{table}
\caption[ ]{Velocity distribution for 1996-1997 telescopic sporadics}
\begin{flushleft}
\begin{tabular}{|l|ccccccc|r|}
\hline
\hline
Year & 0 & A & B & C & D & E & F & Tot. \\
\hline
\hline
1996 & 1 & 1 &  6 & 13 & 19 & 10 &  4 & 54 \\
1997 & 2 & 4 & 11 & 31 & 43 & 30 & 17 & 138 \\
\hline
Tot. & 3 & 5 & 17 & 44 & 62 & 40 & 21 & 192 \\
\hline
\hline
\end{tabular}
\end{flushleft}
\end{table}
\end{center}

\section{Discussion}

We presented here the results of visual and telescopic observations of
$\alpha$-Cygnid stream made by CMW observers in years 1995--1997. This
stream is not included in the IMO List of Visual Meteor Showers (Rendtel
et al 1995) due to the lack of photographic, video and radio data confirming its
presence. The only continuous (not visual) surveys are the Harvard Super
Schmidt Program of the 1952-54 period and the Harvard Radar Project in
the 1960's. Both have gaps in coverage due to weather or instrument
irregularities. Unfortunately the $\alpha$-Cygnids seem to fall in such gaps.
McCrosky and Posen (1961) presented the orbital elements of 2529
meteors photographed simultaneously from two camera stations of the
Harvard Meteor Project. The mean number of meteors captured in periods
January--June and August--December is 221 events per month. The number
of meteors photographed in July is only 102. It is over two times
smaller than in other months and certainly it is the reason of lacking
the meteors from $\alpha$-Cygnid stream in that project.
The similar situation occurred during the Harvard Radar Project.
However recent radio results obtained by Michael
Boschat from Dalhousie University in Canada (Boschat 1998) showed clear
enhancement of radio echoes in days 1998 July 19-20.

Analysis of 2748 paths of meteor events observed in July
1995, 1996 and 1997 allowed us to obtain the basic properties of the
stream. The geocentric velocity of the $\alpha$-Cygnid events
is $V_\infty=41$~km/s and daily drift of
the radiant  $\Delta\alpha=+0.6^\circ$, $\Delta\delta=+0.2^\circ$.
The coordinates the center of the radiant
of $\alpha$-Cygnid stream are $\alpha=302.5^\circ$ and
$\delta=+46.3^\circ$.

We performed similar analysis for 234 telescopic meteors plotted also by CMW
observers. Obtained parameters of the stream are in very good
agreement with results derived from visual data.

Our results are also consistent with visual observations by Denning 
(1919) and Jenniskens (1994) and photographic results obtained by
Babadzhanov \& Kramer (1965).

From magnitude distribution of 738 $\alpha$-Cygnids we obtained the
population index $r$ equal to $2.55\pm0.14$ which is in good agreement
with previously obtained value $r=2.7$ (Jenniskens 1994).

The velocity distributions of 754 $\alpha$-Cygnids and 4339 sporadics are
different with probability higher than 0.999 which gives another proof
for reality of the $\alpha$-Cygnid stream.

From visual observations made by CMW members in years 1995--1997 we
obtained the clear activity profile of the $\alpha$-Cygnid stream.
Meteors belonging to this shower were detected from June 30 to July 31
with clear maximum near July 18 ($\lambda_\odot=116.5^\circ$). The
maximal ZHRs reached the level $3.6\pm1.2$. These results are in very
good agreement with results presented by Jenniskens (1994) who obtained
$\lambda_{\odot{\rm (max)}}=116.0\pm0.5^\circ$ and ${\rm
ZHR_{max}}=2.5\pm0.8$.

In spite of making many photographic exposures we still have no
photographic confirmation of this stream. To confirm of disprove our
results further visual, telescopic and particularly video and
photographic observations are clearly needed.

\begin{acknowledgements} We would like to thank to all observers who sent
us their observations. We are especially
grateful to Prof. Jerzy Madej for helpful discussions, reading and
commenting on the manuscript and also to Dr. Jacek Cho{\l}oniewski for
many helpful hints. This work was supported by KBN
grants 2~P03D~020~11 and 2~P03D~002~15 to A. Olech.
\end{acknowledgements}

\begin{figure}
\picplace{8.5cm}
\caption[ ]{The radiant of $\alpha$-Cygnids resulting from Denning's 
(1919) observations. Assumed parameters are: $V_\infty=37$~km/s,
$\lambda_{\odot(max)}=116^\circ$, $\Delta\alpha=+0.6^\circ$ and
$\Delta\delta=+0.2^\circ$. Number of the events is 20.}
\end{figure}
\includegraphics{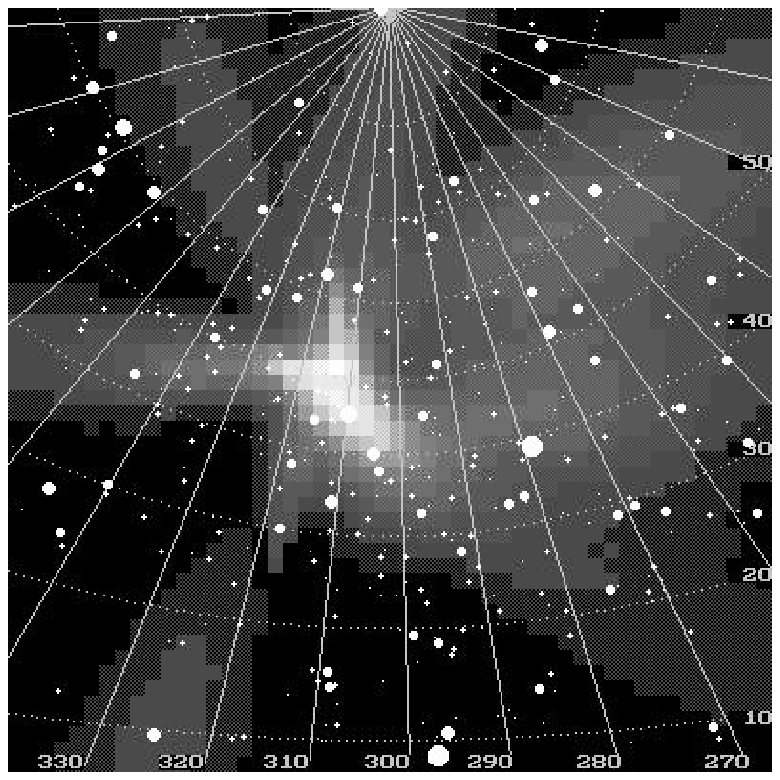}

\begin{figure}
\picplace{8.5cm}
\caption[ ]{The radiant of $\alpha$-Cygnids resulting from CMW visual
data. Assumed parameters are: $V_\infty=41$~km/s,
$\lambda_{\odot(max)}=116^\circ$, $\Delta\alpha=+0.6^\circ$ and
$\Delta\delta=+0.2^\circ$. Number of the events is 2748.}
\end{figure}
\includegraphics{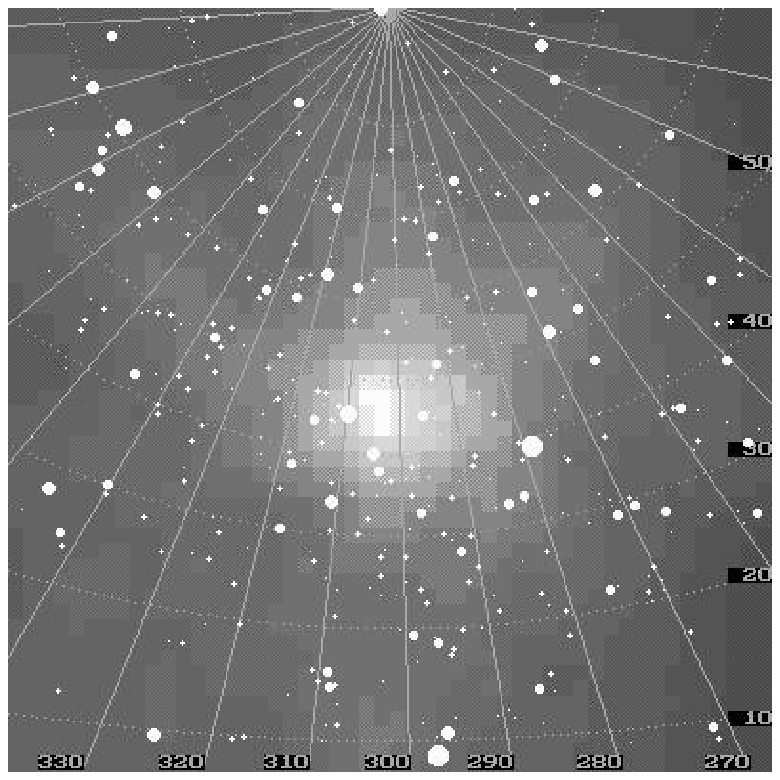}

\begin{figure}
\picplace{8.5cm}
\caption[ ]{The radiant of $\alpha$-Cygnids resulting from CMW
telescopic data. Used parameters are: $V_\infty=40$~km/s,
$\lambda_{\odot(max)}=116^\circ$, $\Delta\alpha=+0.6^\circ$ and
$\Delta\delta=+0.2^\circ$. Number of the events is 234.}
\end{figure}
\includegraphics{ds7893fig.ps}

\newpage
\begin{figure}
\picplace{8.8cm}
\caption[ ]{The activity profile of $\alpha$-Cygnids during 1995--1997.
The solid line represent the fit given in equation (4).}
\end{figure}
\includegraphics{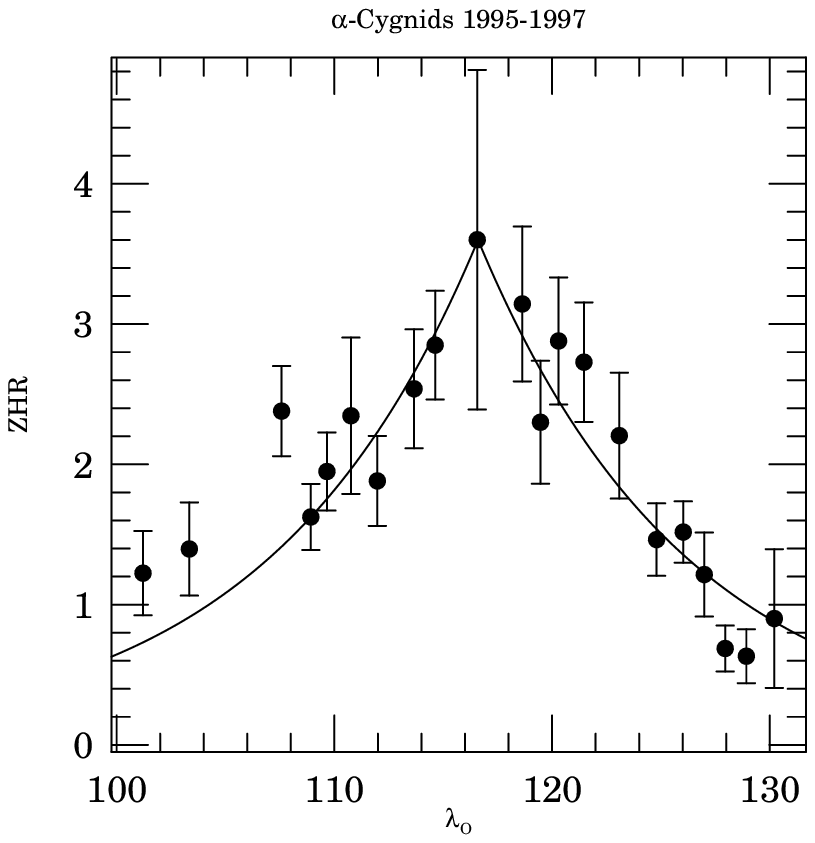}

\end{document}